\documentclass{pic2012}

\usepackage{hyperref}

\newcommand{\pt}{$p_{\mathrm{T}}$}

\usepackage{lineno}

\def\runauthor{Tulin Varol}
\def\shorttitle{Muon Performance in the Presence of High Pile-up in ATLAS}
\begin{document}

\title{MUON PERFORMANCE IN THE PRESENCE OF HIGH PILE-UP IN ATLAS
}

\author{PIC~2012\\
TULIN VAROL\\
On behalf of the ATLAS Collaboration}

\address{Department of Physics\\
University of Massachusetts, Amherst, MA 01002\\
E-mail: tvarol@physics.umass.edu }

\maketitle

\abstracts{In 2012, the LHC is operated at $\sqrt{s}$ = 8 TeV in a mode leading up to 40 inelastic \textit{pp} collisions per bunch crossing. The identification and reconstruction of muons produced in hard collisions is difficult in this challenging environment. Di-muon decays of \textit{Z} bosons have been used to study the muon momentum resolution as well as the muon identification and reconstruction efficiencies of the ATLAS detector as a function of the muon transverse momentum from $p_{\mathrm{T}}$ = 15 GeV to 100 GeV and the number of inelastic collisions per event. These studies show that the muon momentum resolution, muon identification and reconstruction efficiencies are independent of the amount of pile-up present in an event. 
}

\section{Introduction} The ATLAS detector is optimized for muon identification, with an efficiency greater than 95\% and a fractional momentum resolution better than 3\% over a wide transverse momentum (\pt) range and 10\% at \pt\ = 1 TeV. Muons are independently measured in the Inner detector (ID) and the Muon Spectrometer (MS). 

The Inner Detector is designed to provide a precise measurement of charged particle trajectories in a dense track environment. The ID tracks are reconstructed within the pseudorapidity\footnote{The pseudorapidity $\eta$ = - $\ln(\tan(\theta/2))$, where $\theta$ is the polar angle measured from the beam line.} range $ | \eta | < $ 2.5 in a solenoidal field of 2 T. Three detector technologies are exploited, covering different radial distances from the interaction point: a silicon pixel detector, a semiconductor tracker (SCT), and a transition radiation tracker (TRT).
The silicon pixel detector provides precise vertexing by achieving the highest granularity around the vertex region. The SCT allows impact parameter measurements and vertex reconstruction and also ensures an accurate particle momentum measurement. The TRT eases the pattern recognition, contributes to electron identification and it has also a major role in improving the ID track momentum resolution. Each detector consists of barrel and end-cap regions in order to minimize the material traversed by particles coming from the interaction point.

The Muon Spectrometer allows for precise momentum measurements independently of the inner tracking detector and provides identification and reconstruction within the pseudorapidity range $ | \eta | < $  2.7. A barrel toroid and two end-cap toroids produce a toroidal magnetic field of approximately 0.5 T and 1 T for the muon detectors in the central and end-cap regions, respectively. The MS exploits four types of detectors. The deflection of muons by the magnetic field generated by three air core toroidal magnets, one in the barrel and one in each end-cap, is measured by three layers of precision drift tubes (MDT) chambers in  $ | \eta | < $~2 and two layers of the MDT chambers in combination with one layer of cathode strip chambers (CSC) at the entrance of the MS for 2 $ \leq  | \eta | < $ 2.7. The MDT and CSC measure the position in the bending plane, namely $\eta$. CSC also provides the measurements in the non-bending plane by means of CSC $\phi$ strips. The resistive-plate chambers (RPC) in the barrel region~($ | \eta | < $ 1) and the thin-gap chambers (TGC) in the end-caps (1 $ <  | \eta | < $ 2.4) provide triggering and also measure the muon trajectory in the non-bending plane of the spectrometer magnets. Triggering with the RPC and TGC happens thanks to the $\eta$ measurement in those chambers, in addition to the $\phi$ measurement.

Pile-up is defined as the average number of particle interactions per bunch crossing $\mu$, and is directly correlated with the instantaneous luminosity. The growing pile-up presents a challenge for the ATLAS detector and its subsystems during data taking. 

Measurements of muon identification efficiency and muon momentum resolution are presented using $\sqrt{s}$ = 8 TeV data collected in 2012, with up to 40 inelastic \textit{pp} collisions per bunch crossing. The effect of increasing pile-up on the performance due to increasing luminosity is investigated by studying the muon identification efficiency under different pile-up conditions. 


\section{Muon Identification Strategies}
 In ATLAS, four kinds of muon candidates are distinguished depending on the way they are reconstructed:

\textbf{Stand-alone muon:} The spectrometer track is entirely reconstructed in the MS and then extrapolated back to the beam line to determine the direction of flight and the impact parameter of the muon at the interaction point. To obtain the muon momentum at the interaction point, the muon momentum measured in the MS is corrected for the parametrized energy loss in the calorimeter.

\textbf{Combined muon:} Tracks result from the combination of MS and ID measurements by a statistical combination or a refit of the entire track. Energy losses in the calorimeter are taken into account using parametrization and possibly calorimeter measurements. This reconstruction strategy provides the most precise measurement of the momentum and position of a muon.

\textbf{Segment tagged muon:} If a track in the ID, extrapolated to the MS, is associated with straight track segments in the precision muon chambers, it is identified as a muon. This provides efficiency recovery in regions with low MS detector coverage. 

\textbf{Calorimeter tagged muon:} An ID track with an energy deposition in the calorimeters compatible with a minimum ionizing particle is identified as a muon. This is used in place of ID tracks to reduce the background in the Òtag and probeÓ method.
 
  \vspace{1cm}

\section{Muon Reconstruction Efficiency}
The muon reconstruction efficiency based on the tag-and-probe method is determined as a function of $\eta$ and $\mu$. The efficiencies measured with experimental data using muon pairs produced in the decays of \textit{Z} bosons are compared with muon reconstruction efficiencies predicted by the Monte Carlo (MC) simulation. 

\begin{figure}[htb]
  \begin{center}
   \includegraphics[width=0.49\textwidth]{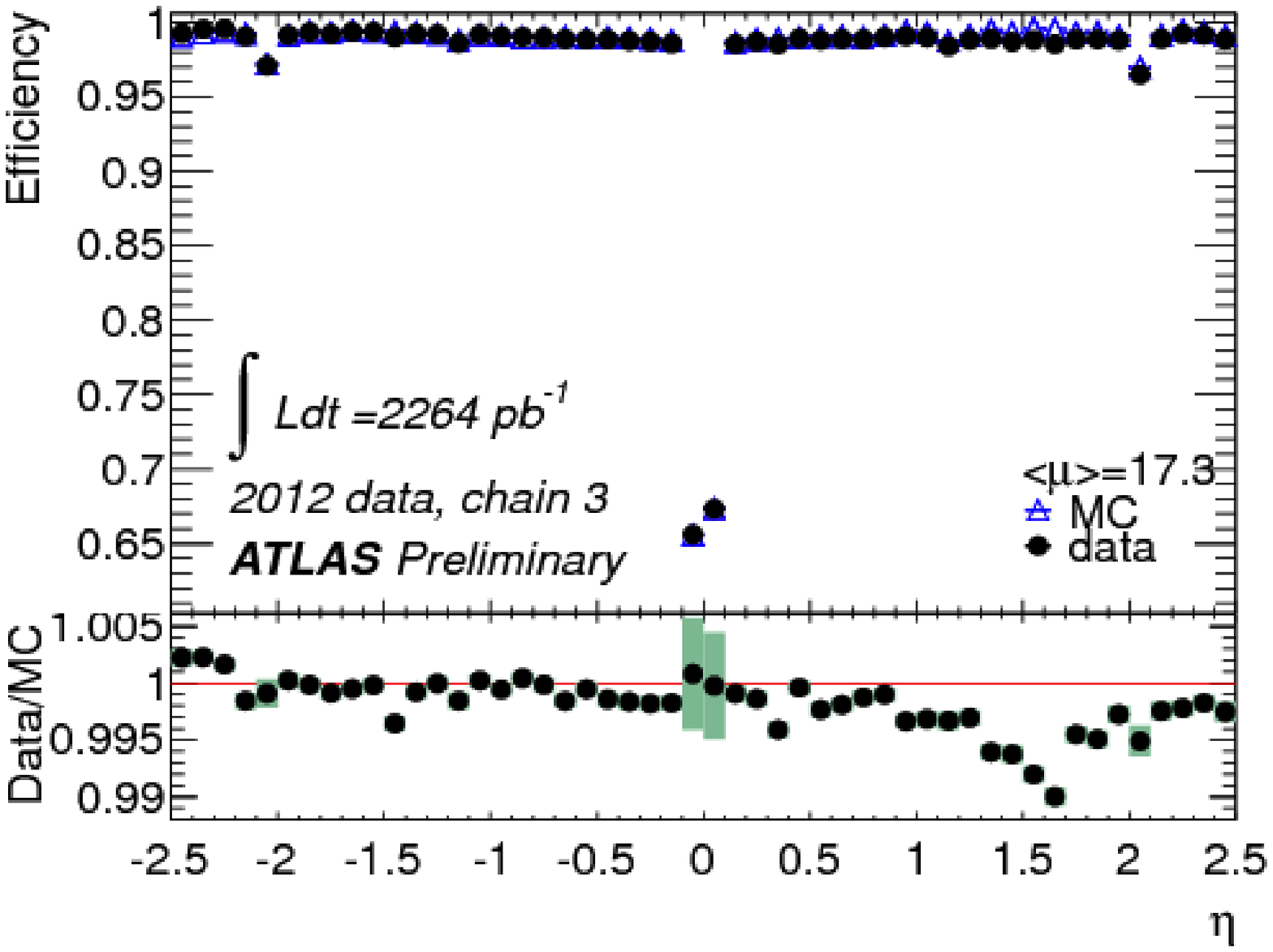}
   \includegraphics[width=0.49\textwidth]{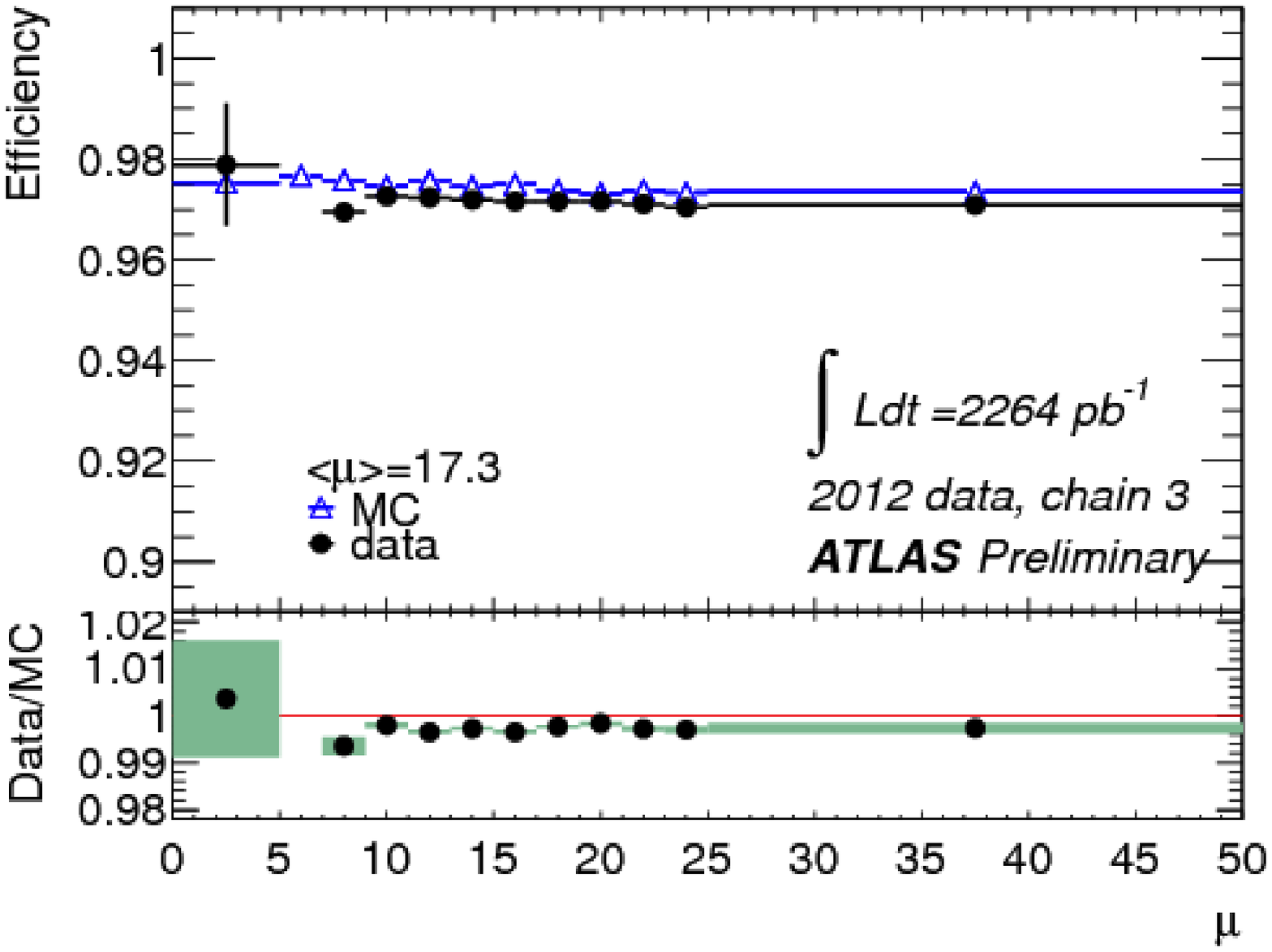}
  \end{center}
  \vspace{-0.6 cm}
  \caption[*]{Muon reconstruction efficiency as a function of $\eta$ (left) and $\mu$ (right) for muons identified with either combined or segment-tagged algorithms \cite{ATL2}.}
  \label{fig:momIdenEff}  
\vspace{-1.4cm}
\end{figure}

Tag-and-probe method is often used in the muon performance studies since it provides a clean sample of muon candidates. In this study, the tag-and-probe method selects events using $\textit{Z}\rightarrow\mu\mu$ decays with one well reconstructed muon, the tag, and one opposite charged track measured by the ID, the probe. To reduce the backgrounds, the invariant mass of the pair has to be close to the nominal \textit{Z} mass to match the tag together with the probe to the signature of a \textit{Z} boson decay. The reconstruction efficiency is determined from the fraction of probes which are matched to a reconstructed muon. The muon reconstruction efficiency~vs~$\eta$ and $\mu$ are shown on the left and right respectively in Fig. \ref{fig:momIdenEff}. The errors shown on the ratios are statistical. On the left plot, the efficiency drop within the pseudo rapidity range 1 $ <  | \eta | < $ 2 is due to the underestimation of the misalignment in the transition region since there is no optical alignment for some chambers. The increasing pile-up shows no significant effect on the efficiency.

  \vspace{-0.6 cm}

\section{Muon Momentum Resolution}
\begin{figure}[htb]
  \begin{center}	
    \includegraphics[width=0.41\textwidth]{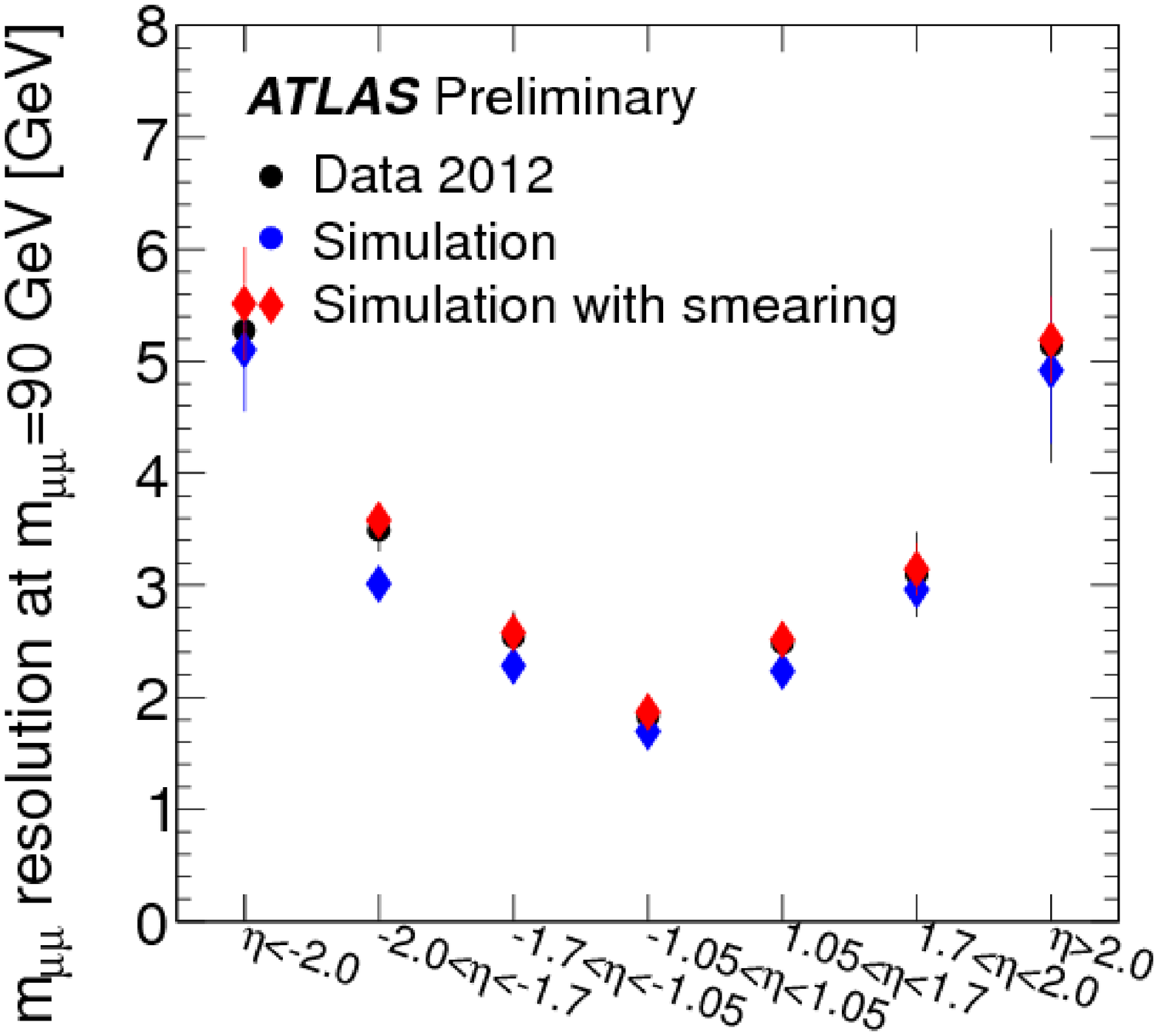}
    \includegraphics[width=0.41\textwidth]{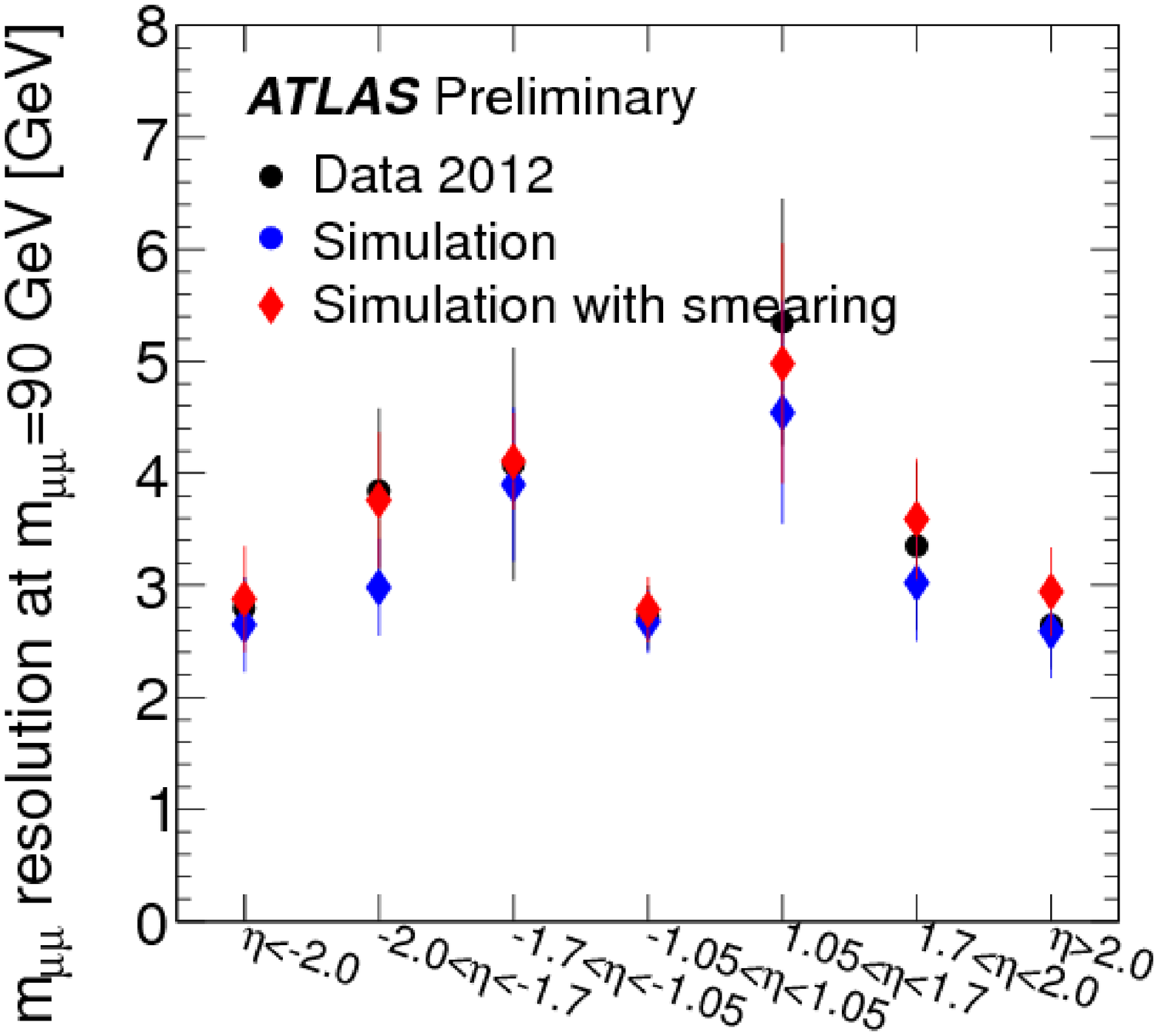}
  \end{center}
   \vspace{-0.5cm}
\caption[*]{Inner Detector di-muon mass resolution (left) and Muon Spectrometer di-muon mass resolution (right) \cite{ATL2}.} 
\label{fig:momRes}
\vspace{-.7cm}
\end{figure}

\begin{figure}[!thb]
 \vspace*{-0.1cm}
 \begin{center}
    \includegraphics[width=0.41\textwidth]{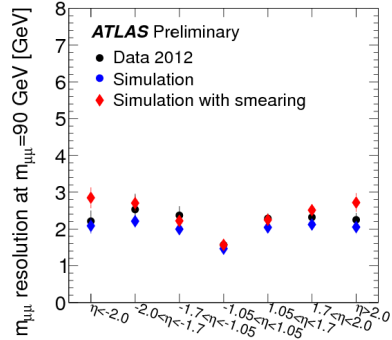}
 \end{center}
 \vspace{-0.6cm}
\caption[*]{Combined di-muon mass resolution near the \textit{Z} peak \cite{ATL2}.}
\label{fig:comRes}
\vspace{-.5cm}
\end{figure}

A measurement of the muon momentum resolution is performed using 824 $pb^{-1}$ of data collected in early 2012 and compared to $\textit{Z}\rightarrow\mu\mu$ simulation sample. The di-muon invariant mass distributions are obtained separately from MS and ID track parameters and integrated over all muon \pt~values. The resolution is the width of the Gaussian which is convoluted with the Breit-Wigner shape in $\textit{Z}\rightarrow\mu\mu$ decays at generator level. The fit is performed in the mass window $m_{\mu\mu} \in$ [75 GeV,~105~GeV]. 
The MS and ID mass resolutions are shown in Fig. \ref{fig:momRes}  as a function of $\eta$ region. The simulation reproduces the data well. In MS resolution, disagreement with simulation is due to the residual MS misalignment. Fig. \ref{fig:comRes} shows the combined di-muon mass resolution. Error bars are the sum of the statistical error and the absolute value of the change of the resolution when the fit range is reduced to $m_{\mu\mu} \in$ [82 GeV, 100 GeV] from $m_{\mu\mu} \in$ [75 GeV, 105 GeV].

\section{Results} 
Muon identification efficiency and muon momentum resolution studies have been carried out with $\sqrt{s}$ = 8 TeV data recorded in 2012. The simulation is observed to describe the data well. The effect of high pile-up is studied by using the reconstruction efficiency. No significant pile-up dependence is observed. 
For 2012, with increasing luminosity, hence pile-up, ATLAS continues to closely test the performance of the muon reconstruction.


\begin{thebibliography}{0}
%
\bibitem{JINST} ATLAS Collaboration, JINST {\bf3 S08003} (2008).
%
\bibitem{ATL2} {\raggedright ATLAS Collaboration, \\ 
\href{https://atlas.web.cern.ch/Atlas/GROUPS/PHYSICS/MUON/PublicPlots/2012/June/index.html}%
{https://atlas.web.cern.ch/Atlas/GROUPS/PHYSICS/MUON/PublicPlots/2012/June/index.html}.}
%
\bibitem{ATL3} {\raggedright ATLAS Collaboration, ATLAS-CONF-2011-046,\\ 
\href{https://cdsweb.cern.ch/record/1338575}{https://cdsweb.cern.ch/record/1338575}.}
%
\bibitem{ATL4} {\raggedright ATLAS Collaboration, ATLAS-CONF-2010-064, \\
\href{http://cdsweb.cern.ch/record/1281339?}{http://cdsweb.cern.ch/record/1281339}.}
%
\bibitem{ATL5} {\raggedright ATLAS Collaboration, ATLAS-CONF-2011-008, \\
\href{http://cdsweb.cern.ch/record/1330715}{http://cdsweb.cern.ch/record/1330715}.}
%
\end{thebibliography}
\end{document}